\documentclass[%
 reprint,
 showkeys,
 amsmath,amssymb,
 aps,
 pra,
 nofootinbib 
]{revtex4-2}

\usepackage{graphicx}
\usepackage{dcolumn}
\usepackage{bm}
\usepackage{color} 
\usepackage{lipsum} 
\usepackage{amsmath} 
\usepackage{amsthm} 
\usepackage{natbib}	
\setcitestyle{numbers} 
\usepackage{braket}
\usepackage{enumitem} 
\usepackage{hyperref}
\hypersetup{
    colorlinks=true,
    urlcolor=blue,
    linkcolor=blue,		
    citecolor=blue}
\usepackage{soul} 

\renewcommand{\theenumi}{\roman{enumi}} 

\usepackage{xcolor}

\begin{document}


\title{Energy additivity as a requirement for
	universal \\ quantum thermodynamical frameworks}


\author{Luis Rodrigo Neves}
\email{rodrigoneves@usp.br}
\affiliation{
 Instituto de F\'isica de S\~ao Carlos, Universidade de S\~ao Paulo -- S\~ao Carlos (SP), Brasil 
}
\author{Frederico Brito}
\email{frederico.brito@tii.ae, fbb@ifsc.usp.br}
\affiliation{
 Instituto de F\'isica de S\~ao Carlos, Universidade de S\~ao Paulo -- S\~ao Carlos (SP), Brasil 
}
\affiliation{Quantum Research Centre, Technology Innovation Institute, P.O. Box 9639, Abu Dhabi,UAE}

\date{\today}

\newcommand{\tr}{       
	\text{tr} 
}
\newcommand{\avg}[1]{       
	\braket{#1}
}
\newcommand{\hc}{^\dagger}

\newcommand{\kb}[1]{    
	\ket{#1}\bra{#1}
}

\newcommand{\kbb}[2]{   
	\ket{#1}\bra{#2}
}

\newcommand{\coupstr}{K} 

\theoremstyle{definition}
\newtheorem{definition}{Definition}
\newtheorem{remark}{Remark}[definition]
\newtheorem{proposition}{Proposition}



\begin{abstract}

	The quest to develop a general framework for thermodynamics, suitable for the regime of strong coupling and correlations between subsystems of an autonomous quantum ``universe,'' has entailed diverging definitions for basic quantities, including internal energy. While most approaches focus solely on the system of interest, we propose that a universal notion of internal energy should also account for the environment in order to keep consistency with the closed-system energy of the universe. 
	We introduce an abstract framework to describe all effective Hamiltonian-based approaches and address a rigorous definition of energy additivity in this context, in both a weak and a strong forms, discussing the underlying subtleties.
	As an illustration, we study a particular two-qubit universe model, obtaining the exact master equations for both parties and calculating their effective Hamiltonians and internal energies as given by the recently devised minimal dissipation approach. In this case, we show that internal energies are neither additive nor conservative, which leads to unphysical features.

\end{abstract}


\maketitle

\section{Introduction}\label{sec:intro}

	Within the contemporary field of quantum thermodynamics \cite{Vinjanampathy_Anders_2016, Alicki_Kosloff_2018,  Millen_Xuereb_2016, Dann_Kosloff_2023, Deffner_Campbell_2019}, the past few years have witnessed an increasing concern on physical setups that largely deviate from the once paradigmatic picture of a small system of interest ($S$) under external driving and weakly coupled to a heat bath ($E$) \cite{Alicki_1979}, shifting the focus towards strong system-environment coupling and correlations, far-from-equilibrium environments, non-Markovian dynamics, and exact master equations, to name just a few examples \cite{ Micadei_etal_2019, Rivas_2020, Nettersheim_etal_2022,Carrega_etal_2016,Bresque_etal_2021, AntoSztrikacs_etal_2023, Rodrigues_Lutz_2024}. 
	
	Notably, several approaches dismiss the classical external agent (work source), in such a way that every thermodynamic phenomenon (e. g. the exchange of work and heat) amounts to the interaction between two parts of a  closed, autonomous, bipartite quantum system, ruled by a constant Hamiltonian operator $H = H^A + H^B + H^I$ \cite{Valente_etal_2018,Huang_Zhang_2022,  Pyharanta_etal_2022, Malavazi_Brito_2022, Alipour_etal_2016,Colla_Breuer_2022, Weimer_etal_2008, Ali_etal_2020, Alipour_etal_2022, Ahmadi_etal_2023, Silva_Angelo_2023}. Here, $H^I$ is the assigned Hamiltonian of interaction and $H^{(j)}$ is the individual one for subsystem $j$. The labels $j=A,B$ are chosen to emphasize that the ``system'' and its ``environment'' are symmetrically conceived, so that the matter of which partition to describe explicitly becomes arbitrary, in principle. We refer to this more general, full-quantum approach as the \textit{autonomous universe} paradigm.
	
	In contrast to the common procedure of relying on specific physical models or assuming severe restrictions on the type of interaction \cite{Carrega_etal_2016, Bresque_etal_2021, AntoSztrikacs_etal_2023, Rodrigues_Lutz_2024, Valente_etal_2018, Huang_Zhang_2022}, the autonomous universe paradigm encompasses many schemes of wider applicability \cite{Weimer_etal_2008, Alipour_etal_2016, Ali_etal_2020, Alipour_etal_2022, Colla_Breuer_2022, Malavazi_Brito_2022, Ahmadi_etal_2023}.
	Indeed, any theoretical framework of intended universal or general reach should hold also in the autonomous paradigm \cite{Weimer_etal_2008, Ali_etal_2020, Alipour_etal_2022, Colla_Breuer_2022}, arguably the border of quantum thermodynamics.
	
	Although the attainability of some form of the First and Second Laws of thermodynamics is frequently highlighted as a validation criterion for new theoretical frameworks, the very definition of the basic quantities involved is still a subject of controversy. In particular, the First Law is often addressed as a matter of how to decompose a change in internal energy into exchanged work and heat, but a physically meaningful definition for the internal energy of an open system should be the point of departure.
	
	Many of the existing proposals define the internal energy $U^{(j)}$ as the mean value of some effective Hamiltonian $\tilde{H}^{(j)}(t)$ \cite{Weimer_etal_2008, Alipour_etal_2016, Colla_Breuer_2022, Pyharanta_etal_2022, Valente_etal_2018, Malavazi_Brito_2022, Huang_Zhang_2022}, embodying part of the interaction energy. Surprisingly enough, just a few of them address the same definitions consistently for the coupled system (``environment'')  \cite{Alipour_etal_2016, Malavazi_Brito_2022}.
	However, in any allegedly universal framework, suitable for the autonomous-universe scenario, the logical counterpart of redefining the energy observable of the system ($A$) is to redefine that of its environment ($B$) \textit{by the same rule} (``what's sauce for the goose is sauce for the gander'').
	
	As the ``universe'' $A\oplus B$ is a closed and isolated system, its total energy is given by a well-known observable, namely the total Hamiltonian $H = H^A+H^B + H^I$, with a constant mean value. 
	In this article, it is put forward that any general, effective-Hamiltonian-based quantum thermodynamical framework, when applied to the autonomous paradigm, should embody these two basic features: (a) the conservation of total internal energy, $U^A + U^B = \text{constant}$; and (b) the additivity of internal energies, $U^A + U^B = \avg{H}$. To highlight their intimate connection, these properties are here named \textit{weak} and \textit{strong additivity},  respectively. By keeping them distinguished, we hope to emphasize both the essential character of weak additivity -- a mere statement of energy conservation within an isolated system -- and some subtleties inherent to the strong form (see Section \ref{subsec:definitions}).

	The general feature concerning the additivity of internal energies emerges when the theoretical picture of a bipartite system is adopted. In this case, either of two classes of physical setups is being modeled: (i) mediated interactions, such as two atoms coupled via an electromagnetic field, and (ii) direct interactions, such as an atom interacting with a quantized electromagnetic mode. In case (i), a complete physical model would explicitly include the mediating field. Instead, the bipartite approximation assumes that the field's influence is entirely captured by the unitary effect of an interaction Hamiltonian $H^I$ acting between the two subsystems. In this framework, any deviation from additivity -- quantified by the mismatch $\Delta :=\avg{H} - U^A - U^B$ -- could only be justified by attributing $\Delta$ to the field, effectively treating it as an external energy reservoir. This, however, violates the assumption of the autonomous universe. Therefore, to maintain consistency, one must either: (a) model the field explicitly, treating the system as tripartite; or (b) refrain from applying thermodynamic reasoning to the reduced bipartite model. A consistent alternative is to introduce a rule that assigns a share of the interaction energy $\avg{ H^I }$ to each subsystem -- effectively the thermodynamic counterpart of maintaining the bipartite universe picture. Under such a rule, internal energies must be defined additively (in the strong sense), or not at all. In case (ii), by contrast, there is no third system to accommodate any energy mismatch. Thus, internal energy additivity becomes a necessary condition for thermodynamic consistency. Notably, many usual models in quantum thermodynamics -- often inherited from the quantum dynamics literature -- fall under class (ii) \cite{Alipour_etal_2016, Colla_Breuer_2022, Valente_etal_2018, Ali_etal_2020, Ahmadi_etal_2023}, although class (i) setups also remain relevant \cite{Micadei_etal_2019, Pal_etal_2020, Nettersheim_etal_2022,   Micadei_etal_2021, Bresque_etal_2021}.

	Furthermore, the additivity of energies has a key role in the limiting case of classical thermodynamics. As is well known, weak additivity ensures the equivalence between maximized entropy and uniform temperature within an isolated compound system
	[\citenum{Callen_1985}, \S 2.4], while strong additivity entails a factorizable canonical partition function, greatly simplifying the marginal statistics [\citenum{Callen_1985}, \S 16.2].	

	Here, a unified formalism is introduced to describe the space of possible effective Hamiltonians, upon which the rigorous notions of additivity are devised.
	 To illustrate the consequences of adopting non-additive prescriptions, we take the minimal dissipation approach \cite{Colla_Breuer_2022}. Among several available effective-Hamiltonian-based theoretical frameworks that do not address additivity (and thus should not be expected to satisfy it), we have chosen this one for the following reasons. First, it is model-independent and thus of very wide applicability in principle. Second, it is the thermodynamical unfolding of the very recent, well-succeeded endeavor of addressing a unique, physically-founded realization of the celebrated Lindblad-form quantum master equations in the generalized realm of non-Markovian (and model-independent) dynamics \cite{Hayden_Sorce_2022}. We should note that, under less rigorous or more specific settings, the idea of defining internal energy from the  ``renormalized'' unitary part of the open-system dynamics has long been employed as the basis of approaches to autonomous-system quantum thermodynamics \cite{Weimer_etal_2008, Valente_etal_2018}. There is also an independent, recent work in which an equivalent definition is adopted for a wide class of environment models \cite{Huang_Zhang_2022}. In summary, the approach is appealing and well suited to existing trends in the literature. 
	
	The task of probing additivity for given definitions of quantum internal energy is typically toughened under the paradigm of an infinite-dimensional environment. To avoid the technical difficulties, we take both ``system'' $A$ and ``environment'' $B$ as two-level systems. Several settings of this nature have been studied in quantum thermodynamics, both theoretically and in experimental setups \cite{Jevtic_etal_2012, Micadei_etal_2019, Pal_etal_2020,  Micadei_etal_2021, Bresque_etal_2021}. In the more typical case of $B$ as a harmonic mode, this could be regarded as an effective model for the low-lying energy states dynamics. Under this assumption, and with a particularly simple interaction Hamiltonian, we are able to obtain fully analytic expressions for the internal energy as given by the canonical Hamiltonian of \cite{Hayden_Sorce_2022}. Then, as this article's main result, it is shown that the minimal-dissipation internal energy does not satisfy any form of additivity, that is, it predicts a time-dependent effective total energy for a closed, autonomous, bipartite universe.
	
	The paper is outlined as follows.
	In Section \ref{sec:defs_addit_quant}, definitions of weak and strong additivity of internal energies are introduced, in an abstract setting which should encompass all effective-Hamiltonian-based approaches; connection with previous works is explicitly addressed in \ref{sec:connec_lit}.
	In Section \ref{sec:canon_deph} we review the approach of \cite{Colla_Breuer_2022} under our abstract framework. 
	Besides deriving a no-go result from elementary features of this approach, we apply it to a particular two-qubit model, obtain the exact master equations, and discuss its unusual predictions. 
	The conclusions are summarized in Section \ref{sec:conclusions}.

\section{Scope and definitions}
\label{sec:defs_addit_quant}

\subsection{Framework}
\label{subsec:framework}

	Let the Hilbert space $\mathcal{H}=\mathcal{H}^A\otimes\mathcal{H}^B$ describe a closed, bipartite quantum system (henceforth ``universe''), where $\mathcal{H}^{(j)}$ are the subsystems' Hilbert spaces. As our focus here is on an autonomous universe, we let its Hamiltonian $H$ be constant.
	We denote by $\mathcal{B}\left(\mathcal{A}\right)$ the space of all Hermitian operators on an arbitrary Hilbert space $\mathcal{A}$, and by $\mathcal{S}\left(\mathcal{A}\right)\subset \mathcal{B}\left(\mathcal{A}\right)$ the subset of all allowed quantum states (i. e., non-negative, unit-trace, Hermitian operators). Within the autonomous universe paradigm, any definition of effective Hamiltonians for $A,B$ should unambiguously assign a pair of local observables given the initial preparation of the universe, the (constant) total Hamiltonian, and a point in time, suggesting the following formalization.
	
	    \begin{definition}\label{def:eff_ham_rule}
		An effective Hamiltonian rule (EHR) $\mathbf{E}$ is a correspondence of the type
		
		\begin{equation}
			\mathbf{E}\left(
			\rho_0, H, t
			\right) = 
			\left(
			\tilde{H}^A(t),
			\tilde{H}^B(t)
			\right),
		\end{equation}
		i. e., it is a mapping 
		$\mathbf{E}: D\left(\mathbf{E}\right) 
		\rightarrow
		\mathcal{B}\left(\mathcal{H}^A \right)\times\mathcal{B}\left(\mathcal{H}^B \right)$, where the domain $D(\mathbf{E})$ is a subset of $\mathcal{S}\left( \mathcal{H} \right) \times \mathcal{B}\left( \mathcal{H} \right) \times \left[0,\infty\right)$
		satisfying the requirement below.
	\end{definition}
	
	\begin{remark}\label{remarkdef:EHR_domain}
		If $(\rho_0,H,t)\in D\left(\mathbf{E}\right)$, then $(\rho_0,H,0)\in D\left(\mathbf{E}\right)$.
	\end{remark}

We allow subsets in order to encompass common restrictions, as some approaches allow only product \cite{Huang_Zhang_2022, Colla_Breuer_2022} or pure \cite{Valente_etal_2018, Malavazi_Brito_2022} states, only certain classes of interaction models \cite{Valente_etal_2018, Huang_Zhang_2022}, or perhaps an evolution limited in time. 
The remark \ref{remarkdef:EHR_domain} ensures that $t=0$ is an available reference point in time.
Then the several effective Hamiltonian definitions found in 
\cite{Weimer_etal_2008, Alipour_etal_2016, Colla_Breuer_2022, Pyharanta_etal_2022, Valente_etal_2018, Malavazi_Brito_2022, Huang_Zhang_2022} all fit within this abstract structure (up to privileged local bases in
\cite{Weimer_etal_2008} or an \textit{ad hoc} constant in \cite{Alipour_etal_2016, Pyharanta_etal_2022}), insofar as only time-independent Hamiltonians are considered. Here it should be noted that any $H\in\mathcal{B}\left(\mathcal{H}\right)$ admits the usual decomposition $H = H^A+H^B+H^I$, which is essentially unique (see Appendix \ref{sec:app:decomp_H}). 

Whereas several properties are usually expected to be satisfied by the rule $\mathbf{E}$, here we focus on additivity.

\subsection{Energy additivity}
\label{subsec:definitions}

\begin{definition}[\textit{Conservation law}]	
	\label{def:weak_add}
	An effective Hamiltonian rule $\mathbf{E}$ is \textit{weakly additive} if, for any $( \rho_0,H, t)$ in $D\left( \mathbf{E} \right) $, the following holds:
	
	\begin{equation}
		\label{eq:addit_quant:weak_addit_def}
		\avg{
			\tilde{H}^A(t)
		}
		+ 
		\avg{ 
			\tilde{H}^B(t)
		}
		= 
		\avg{
			\tilde{H}^A(0)
		}
		+ 
		\avg{ 
			\tilde{H}^B(0)
		}.
	\end{equation}
	
\end{definition}

Here, $\avg{\mathcal{O}}$ is the usual quantum-mechanical average $\tr\left(\rho(t') \mathcal{O} \right)$ evaluated at the instantaneous state $\rho(t')=e^{-iHt'}\rho_0e^{iHt'}$. In this work we take $\hbar=1$.

In order to define strong additivity, we would like to replace the right hand side of \eqref{eq:addit_quant:weak_addit_def} with the universe internal energy:

\begin{equation}
	\label{eq:addit_quant:strong_addit_def}
	\avg{
		\tilde{H}^A(t)
	}
	+ 
	\avg{ 
		\tilde{H}^B(t)
	}
	= 
	\avg{H}.
\end{equation}

Nevertheless, $\avg{H}$ is not a physical quantity if we consider all Hamiltonian operators in the family  $\left\lbrace H_{\alpha} = H_0+\alpha \mathbb{I}, \alpha\in\mathbb{R}\right\rbrace$ as equivalent, which would be indisputable insofar as only the dynamical effects of $H$ were concerned. In other words, the energy observable is often defined only up to additive constant and, to this extent, the average internal energy of the universe per se is not a well-defined physical quantity. Then, two alternative approaches could be employed to make sense of strong additivity:

{
	\renewcommand{\theenumi}{\alph{enumi}} 
	\begin{enumerate}
		\item Strong additivity could be defined as an invariant property, i. e., a strongly additive rule $\mathbf{E}$ should be ``sensitive'' to any transformation of the type $H\mapsto H + \alpha\mathbb{I}$ within its domain, preserving \eqref{eq:addit_quant:strong_addit_def};
		\label{item:invariant}
		\item A convention could be introduced to fix the identity term in $H$, so that every family $H_{\alpha} = H_0+\alpha \mathbb{I}$  would admit a single ``physically correct'' choice for $\alpha$. 
		\label{item:selective}
	\end{enumerate}
}
Each perspective leads to a corresponding notion of strong additivity:

\begin{definition}
	\label{def:str_add_inv}
	An effective Hamiltonian rule $\mathbf{E}$ is \textit{strongly additive by invariance} if both are true:
	
	\begin{enumerate}
		\item its domain $D\left( \mathbf{E} \right)$ is closed under uniform shift of the Hamiltonian, i. e.,
		
		\begin{equation}
			\begin{gathered}
				\left( \rho_0,H, t \right) 
				\in 
				D\left( \mathbf{E} \right) 
				\Rightarrow \\
				\left( \rho_0,
				H + \alpha\mathbb{I}, 
				t \right)
				\in 
				D\left( \mathbf{E} \right) ,
				\forall \alpha \in \mathbb{R};
			\end{gathered}
		\end{equation}
		\item for any $( \rho_0,H, t) \in D\left( \mathbf{E} \right)$, equation \eqref{eq:addit_quant:strong_addit_def} holds.
	\end{enumerate}
\end{definition}

\begin{definition}
	\label{def:str_add_select}
	An effective Hamiltonian rule $\mathbf{E}$ is \textit{strongly additive by selectivity} if both are true:
	
	\begin{enumerate}
		\item its domain $D\left( \mathbf{E} \right)$ does not admit any uniform shift of the Hamiltonian, i. e.,
		
		\begin{equation}
			\begin{gathered}
				\left( \rho_0,H, t \right) 
				\in 
				D\left( \mathbf{E} \right) 
				\Rightarrow  \\
				\left( \rho_0,
				H + \alpha\mathbb{I}, 
				t \right)
				\notin 
				D\left( \mathbf{E} \right) ,
				\forall \alpha \neq 0;
			\end{gathered}
		\end{equation}
		\item for any $( \rho_0,H, t) \in D\left( \mathbf{E} \right)$, equation \eqref{eq:addit_quant:strong_addit_def} holds.
	\end{enumerate}
\end{definition}

The drawback of Definition \ref{def:str_add_inv} is that such effective Hamiltonians would have to depend explicitly on the global Hamiltonian $H$, to an extent that even a uniform shift of the latter would have an influence on the former. In particular, approaches deriving an effective Hamiltonian directly from the reduced-state dynamical equation \cite{Colla_Breuer_2022,Huang_Zhang_2022} or its trajectory \cite{Weimer_etal_2008, Valente_etal_2018}
could not satisfy this form of additivity, as we will see further for the former case (Definition \ref{def:dynam_reduc}).

In turn, Definition \ref{def:str_add_select} carries the notion that only one member of each family $\left\lbrace H_{\alpha} = H_0+\alpha \mathbb{I}, \alpha\in\mathbb{R}\right\rbrace$ is allowed. The implicit idea is precisely the existence of a certain rule eliminating the arbitrary term of the Hamiltonian, as speculated above in (\ref{item:selective}). Although we do not often see explicit accounts of that matter on the literature, two existing practices can be identified and might be systematized into rules: 

\begin{enumerate}
	\item \label{item:selective_traceless}Impose $\tr H = 0$. This is equivalent to an imposition on the reduced-system Hamiltonians: $\tr H^A = \tr H^B =0$ (see Appendix \ref{sec:app:decomp_H}), and is adopted explicitly, for instance, in \cite{Hayden_Sorce_2022}, in which the approach of \cite{Colla_Breuer_2022} is based. However, this procedure is well-defined only for finite-dimensional quantum systems, since otherwise $H$ could be unbounded, $\tr H=\infty$. Special treatment would then have to be addressed for some important cases, such as a spin-mode models.
	\item \label{item:selective_groundzero}Set each $H^{(j)}$ to have zero ground-state energy. That fixes $H$ as a whole, since the identity ambiguity lies only on the ``local'' part (recall again Appendix \ref{sec:app:decomp_H}). That is the practice observed in many references, in particular whenever the local Hamiltonian is taken to be \textit{proportional} to the excitation number observable, without an offset (see for instance  \cite{Huang_Zhang_2022}). 
\end{enumerate}

It is then straightforward to rigorously define a weak form of internal-energy additivity in the autonomous quantum universe realm. When it comes to the strong form, care must be taken due to the usually accepted indefiniteness of the Hamiltonian, but still additivity could be defined as an invariant property, or an external rule to fix the ground-state energy could be adopted. 

\section{Connection with the literature}

\label{sec:connec_lit}

	Although effective Hamiltonians underlying internal energy are widespread in the autonomous quantum thermodynamics literature 
	 \cite{Weimer_etal_2008, Alipour_etal_2016, Colla_Breuer_2022, Pyharanta_etal_2022, Valente_etal_2018, Malavazi_Brito_2022, Huang_Zhang_2022}, the ``bare'' Hamiltonians alone are sometimes preferred \cite{Ali_etal_2020, Alipour_etal_2022, Ahmadi_etal_2023}. 
	Here, the effective Hamiltonian paradigm is endorsed due to the following: (i) it is empirically well-known that the coupling with an environment redefines the energy levels of a quantum system \cite{Welton_1948,Weiss_2008,colla2025observing}; (ii) it is a common practice to identify a unitary component in the dynamical equation of an open system \cite{Breuer_Petruccione_2002, Rivas_Huelga_2012, de_Vega_Alonso_2017, Zhang_2019, Alipour_etal_2020, AntoSztrikacs_etal_2023}; (iii) it has been shown that a system strongly coupled to a heat bath may reach a steady state that follows the Boltzmann-Gibbs distribution only with respect to a renormalized Hamiltonian \cite{Huang_Zhang_2022}. Regarding (ii), it should be noted that the decomposition of an open-system dynamical equation into unitary and dissipative parts is not unique in general, which was actually the central topic of \cite{Hayden_Sorce_2022}.

	Regardless of the particular definition,  an effective Hamiltonian is meant to incorporate effects from the system-environment (or $A{-}B$) interaction, being thus time-dependent in general, even if the bare Hamiltonian is not. In classical stochastic thermodynamics, important discussions took place regarding whether or not the interaction energy should be computed within the internal energy of a system (inclusive vs. exclusive approaches) \cite{Horowitz_Jarzynski_2008, Peliti_2008b, Vilar_Rubi_2008b}. It turns out that the asymmetric role of an external agent allows for an appropriate treatment of this share of energy, as long as definitions are kept consistent \cite{Peliti_2008a}. Clearly, in autonomous-universe quantum thermodynamics, the inbuilt symmetry between interacting parts makes the matter of how to deal with the (strong) energy of interaction far more subtle. Still, we may identify a correspondence between effective-Hamiltonian-based definitions of quantum internal energy and the inclusive point of view for the classical case, with the difference that, in the former, not all of the interaction energy is ``included'' in the system's account. Accordingly, some authors describe the effective-Hamiltonian approach to internal energy as a ``splitting'' or ``partitioning'' of the interaction energy among the two parts \cite{Alipour_etal_2016, Dou_etal_2018, Huang_Zhang_2022, Seegebrecht_Schilling_2024}.  Strong additivity may be seen as a logical unfolding of such a standpoint. 
	
	Among all existing effective-Hamiltonian approaches to autonomous universe quantum thermodynamics, we are aware of two explicitly addressing both interacting parties and featuring some form of energy additivity by construction. 
	
	The model-independent approach of ref. \cite{Alipour_etal_2016} introduces the idea of a \textit{correlation energy} term, whose average value, when added to the effective internal energies of both individual systems, equals the average universe Hamiltonian. Such a notion of additivity does not fit the definitions proposed here, as it relies on a third share of energy, not assigned to either party. Beyond the intricate implications of dealing with a third entity in the energy balance, this theoretical framework depends on a free parameter for whose determination no general protocol seems to exist. In turn, the Schmidt-decomposition approach of ref. \cite{Malavazi_Brito_2022}, also suitable for any interaction model and strength, entails what was here named invariant strong additivity (Definition \ref{def:str_add_inv}). As we discussed, such an invariance necessarily comes at the expense of some global information built in the definitions of local shares of energy -- in this case, via the choice of a correct phase gauge. As its sole important restriction, this approach assumes the initial state of the universe to be pure. 

Reference \cite{Valente_etal_2018} introduces an effective Hamiltonian for a specific system-environment model, addressing explicit calculations only for the system variables. It can be verified, as shown for a qubit-qubit analog in ref. \cite{Neves_Brito_2023}, that such definition indeed meets additivity under the specific interaction model and initial conditions assumed in \cite{Valente_etal_2018}, whereas such property breaks down under more general settings. According to the concepts addressed here, the corresponding EHR could be considered strongly additive only within a very limited domain.

Finally, ref. \cite{Colla_Breuer_2022} addresses effective Hamiltonians for a universe initially prepared in a product state. The following Section is devoted to this particular EHR and the question of whether it satisfies some form of additivity, as justified in the Introduction.

\section{Example of non-additive internal energies in a simple model}
\label{sec:canon_deph}

\subsection{The ``minimal dissipation'' effective Hamiltonian rule}
\label{subsec:min_dissip_approach}	

	The goal of this section is to apply the approach presented in \cite{Hayden_Sorce_2022, Colla_Breuer_2022} to the two interacting parts of an autonomous universe,  in order to further illustrate the concepts of Section \ref{sec:defs_addit_quant}, especially additivity and its breakdown. We start from a short review of the approach. 
	
	We keep the context defined in Section \ref{subsec:framework} and further assume a factorized initial state, $\rho(0) = \varrho^A(0)\otimes\varrho^B(0)$, whence the state at $t>0$ is $\rho(t) = e^{-iHt} \varrho^A(0)\otimes\varrho^B(0) e^{iHt}$. The reduced state of subsystem $A$ is then $\varrho^A(t) = \tr_B\left[
	e^{-iHt} \varrho^A(0)\otimes\varrho^B(0) e^{iHt}
	\right]$, and we can write 
	
		\begin{equation}
		\label{eq:canon_deph:explicit_map_generic}
		\begin{gathered}
			\varrho^A(t)  = \Phi_t^A\left[\varrho^A(0)\right],\\
			\Phi_t^A := \tr_B\left[
				 e^{-iHt} \cdot \otimes\varrho^B(0) e^{iHt}
			\right],
		\end{gathered}
		\end{equation} 
	 where the map $\Phi_t^A$ is linear by construction. A master equation should assign the reduced state's time derivative, $\dot{\varrho}^A(t)$, as a function of the state itself. By differentiation one gets $\dot{\varrho}^A(t)  = \dot{\Phi}_t^A\left[\varrho^A(0)\right]$ and, if the map is \textit{invertible},\footnote{
		The assumption of an invertible $\Phi_t^A$ is described as ``very weak'' in \cite{Colla_Breuer_2022}. Indeed, if $\mathcal{H}$ is finite-dimensional, the map is an analytic function of time, whence we can be sure that the inverse exists except for isolated points in time; in particular, it exists for some interval $t\in\left[ 0,t_1\right)$ \cite{Hayden_Sorce_2022}.
	} then $\varrho^A(0)  = \left( \Phi_t^A\right)^{-1}\left[\varrho^A(t)\right]$ and we can write
	
		\begin{equation}
		\label{eq:canon_deph:master_eqn_abstract}
			\dot{\varrho}^A(t) = \mathcal{L}^A_t \left[ \varrho^A(t)\right]
		\end{equation}
	where $\mathcal{L}^A_t =\dot{\Phi}_t^A \circ \left( \Phi_t^A\right)^{-1}$. Equation \eqref{eq:canon_deph:explicit_map_generic} shows that the map $\Phi_t^A$  depends only on the universe Hamiltonian $H$, the initial ``environment'' state $\varrho^B(0)$, and time; the same holds for $\mathcal{L}_t^A$.
	
	As discussed in \cite{Hayden_Sorce_2022}, the superoperator $\mathcal{L}^A_t$ can always be written in the usual form 
	
		\begin{equation}\label{eq:canon_deph:gen_master_eq_decomp}
			\mathcal{L}^A_t = 
			-i \left[ \tilde{H}^A(t), \cdot \right] + 
			\mathcal{D}^A_t,
		\end{equation}
	where $\tilde{H}^A(t)$ is Hermitian and the dissipative term can be cast in Lindblad form,
	
		\begin{equation}\label{eq:canon_deph:gen_dissip_lindb}
			\mathcal{D}^A_t =
			\sum_{k}\gamma_k\left(
				L_k \cdot L_k\hc
				-\frac{1}{2}
				\left\lbrace
					L_k\hc L_k, \cdot 
				\right\rbrace 
			\right) .
		\end{equation}
	
	In \eqref{eq:canon_deph:gen_dissip_lindb}, the coefficients $\gamma_k\in\mathbb{R}$ and the jump operators $L_k$ are generally time-dependent. It was shown in \cite{Hayden_Sorce_2022} that, given the superoperator $\mathcal{L}^A_t$ (which is hermiticity-preserving and trace-annihilating, HPTA), there always exists a single dissipator $\mathcal{D}^A_t$ satisfying equations (\ref{eq:canon_deph:gen_master_eq_decomp}, \ref{eq:canon_deph:gen_dissip_lindb}) with $\tr L_k=0$ for every $k$. It was also shown that, if $\mathcal{H}^A$ is finite-dimensional, the traceless choice corresponds to minimizing the dissipative term with respect to a certain norm, whence to name this convention a ``minimal dissipation principle''. Such procedure then fixes the decomposition \eqref{eq:canon_deph:gen_master_eq_decomp}, automatically specifying the effective Hamiltonian $\tilde{H}^A(t)$ up to an identity term. The latter ambiguity is eliminated in \cite{Hayden_Sorce_2022} by assuming a traceless $\tilde{H}^A(t)$ (a well-defined procedure since $\dim{\mathcal{H}^A}<\infty$). This so-called canonical Hamiltonian is the heart of the quantum thermodynamical framework addressed in \cite{Colla_Breuer_2022}, by means of the internal energy scalar
		
		\begin{equation}
			U^A(t) = \avg{ 
				\tilde{H}^A(t)
		 	} 
	 		= \tr{\left[
				\varrho^A(t)\tilde{H}^A(t)
 			\right]}.
		\end{equation}
	
	If the subsystem $B$, usually named an ``environment'', is also finite-dimensional, then every single consideration previously made for $A$ also holds for $B$. In particular, it can be assigned a canonical Hamiltonian $\tilde{H}^B(t)$ and an internal energy $U^B(t) = \avg{\tilde{H}^B(t)}$ following the same definitions. 

	Given  an initial state $\rho_0$ within the subset of product states in $\mathcal{S}(\mathcal{H})$ and a total Hamiltonian $H \in \mathcal{B}(\mathcal{H})$, the construction above unambiguously assigns an effective Hamiltonian to each subsystem at each $t\geqslant 0$.\footnote{
		Provided $\dim \mathcal{H}<\infty$ and that the inverse of $\Phi^{(j)}_t$ exists for both $j$. The latter could be formalized as an extra restriction on the domain $D(\mathbf{E})$, e. g. excluding some times $t$ for certain initial states $\rho_0$, as per the discussion on Appendix \ref{subsec:appx_obtain_eqns_exist}.
	} We are then dealing with an EHR (Definition \ref{def:eff_ham_rule}), which moreover belongs to a special class that we now characterize.

	\begin{definition}\label{def:dynam_reduc}
		An EHR $\mathbf{E}$ is \textit{dynamics-reducible} if both are true:
		\begin{enumerate}
			\item its domain $D(\mathbf{E})$ only admits factorized initial states, $\rho_0=\varrho_0^A\otimes\varrho_0^B$;
			\label{item:dynam_reduc_domain_restriction}
			\item the assignment rule $\mathbf{E}$ can be reduced to functions of the dynamical maps $\Phi_t^{(j)}$, i. e., there exists a proper function $\mathcal{F}$ such that
			\begin{equation}
			\label{eq:canon_deph:dynamics_reducible_EHR}
				\mathbf{E}\left(
				\rho_0, H, t
				\right) = 
				\left(
					\mathcal{F}\left[
						\Phi_t^A
					\right],
					\mathcal{F}\left[
					\Phi_t^B
					\right]
				\right)
			\end{equation}
			for any $\left(
			\rho_0, H, t
			\right)\in D\left(\mathbf{E}\right)$. Note that, in the right side of \eqref{eq:canon_deph:dynamics_reducible_EHR}, each $\Phi_t^{(j)}$ is a function of $\left(
			\rho_0, H, t
			\right)$, as per eq. \eqref{eq:canon_deph:explicit_map_generic} (and its counterpart for $j=B$).
		\end{enumerate}
	\end{definition}

	\begin{proposition}\label{prop:nogo_dynam_reduc}
		An EHR cannot be, simultaneously, strongly additive by invariance (Definition \ref{def:str_add_inv}) and dynamics-reducible (Definition \ref{def:dynam_reduc}).
	\end{proposition}

	\begin{proof}
		Let the EHR $\mathbf{E}$ be strongly additive by invariance, and assume $D\left(\mathbf{E}\right)$ to satisfy the requirement \ref{item:dynam_reduc_domain_restriction} of Definition \ref{def:dynam_reduc}. Let $\mathbf{x}_1=\left(\rho_0,H,t\right)\in D\left(\mathbf{E}\right)$ and take a real $\alpha\neq0$. By Def. \ref{def:str_add_inv} we also have $\mathbf{x}_2=\left(\rho_0,H+\alpha\mathbb{I},t\right)\in D\left(\mathbf{E}\right)$. For simplicity, we represent the corresponding outputs as $\mathbf{E}\left(\mathbf{x}_k\right) = \left( \tilde{H}^A_k, \tilde{H}^B_k \right)$ for $k=1,2$. Then, Def. \ref{def:str_add_inv} also implies that
		
		\begin{align}
		\avg{
		\tilde{H}^A_1
		}
		+ 
		\avg{ 
			\tilde{H}^B_1
		}
		&= 
		\avg{H},
		\label{eq:canon_deph:prop1_avg_sum1}
		\\
		\avg{
			\tilde{H}^A_2
		}
		+ 
		\avg{ 
			\tilde{H}^B_2
		}
		&= 
		\avg{H+\alpha \mathbb{I}}\nonumber\\
		&= \avg{H}+\alpha.
		\label{eq:canon_deph:prop1_avg_sum2}
		\end{align}
	
	However, as the only difference between the $\mathbf{x}_k$ is an identity shift in $H$, eq. \eqref{eq:canon_deph:explicit_map_generic} (and the corresponding for $j=B$) shows that they induce exactly the same dynamical maps $\Phi^{(j)}_t$. Therefore, by further supposing $\mathbf{E}$ to satisfy Def. \ref{def:dynam_reduc}, one is forced to conclude that
	
		\begin{equation}
			\tilde{H}^{(j)}_1=\tilde{H}^{(j)}_2
		\end{equation}
	for both $j$. The latter implies $\avg{
		\tilde{H}^A_1
	}
	+ 
	\avg{ 
		\tilde{H}^B_1
	}=\avg{
	\tilde{H}^A_2
	}
	+ 
	\avg{ 
	\tilde{H}^B_2
	}$, in contradiction with \eqref{eq:canon_deph:prop1_avg_sum1} and \eqref{eq:canon_deph:prop1_avg_sum2}. Hence, Definitions \ref{def:str_add_inv} and \ref{def:dynam_reduc} are incompatible. 
	\end{proof}

	A subtle corollary of the result above is that, if an EHR is dynamics-reducible, then its only possibility of being additive in the strong sense is by means of Definition \ref{def:str_add_select}, implying that it should come along with a convention to fix the trace of $H$, as discussed in Section \ref{sec:defs_addit_quant}. 
	
	In particular, the minimal dissipation EHR \cite{Colla_Breuer_2022}, outlined above, cannot be strongly additive by invariance. As the original reference \cite{Hayden_Sorce_2022} assumes traceless effective Hamiltonians, we here take the convention $\tr{H}=0$ and turn to inquiry whether this EHR is strongly additive by selectivity. We then resort to a particular model.

\subsection{Model: two qubits under a commuting interaction}
\label{subsec:model_two_qubit_deph}

	To allow a fully symmetric and analytic account of the definitions above in a concrete example, as justified in the Introduction, we consider the minimalist choice in which both $A$ and $B$ are two-level systems.
	Each subsystem $j$ has bare Hamiltonian $H^{(j)}$ with eigenstates $\left( \ket{0^{(j)}}, \ket{1^{(j)}}\right)$ (resp. ground and excited), composing a basis to be adopted \textit{in this order}.  As indicated previously, we take $\tr{H^{(j)}}=0$. The bare energy gaps are $\omega^{(j)}$ and therefore $H^{(j)} =
	\left( \omega^{(j)}/2 \right)
	\left( -\ket{0^{(j)}}\bra{0^{(j)}} + \ket{1^{(j)}} \bra{1^{(j)}} \right) =
	\left( \omega^{(j)}/2 \right) \sigma_z^{(j)}$. Note that, in correspondence with the chosen basis ordering, we represent the Pauli matrices as in equation \eqref{eq:solving_dynamics:pauli}. 
	
	The universe Hamiltonian is then
	
		\begin{equation}
		\label{eq:canon_deph:h_tot_def}
		\begin{aligned}
			H&=
			H^A\otimes\mathbb{I}^B+\mathbb{I}^A\otimes H^B+H^I\\
			&= H^L+H^I,
		\end{aligned}
		\end{equation}
	where $H^L:=H^A\otimes\mathbb{I}^B+\mathbb{I}^A\otimes H^B$ denotes the joint local part. We take a commuting interaction term,
	
		\begin{equation}\label{eq:canon_deph:hint_deph}
			H^I = \coupstr \sigma_z^A\otimes\sigma_z^B,
		\end{equation}
	with arbitrary coupling strength $\coupstr\in\mathbb{R}$. This is analogous to the celebrated (pure) dephasing model of qubit-field coupling (see \cite{palma1996, reina2002} and [\citenum{Breuer_Petruccione_2002}, \S 4.2]),
	as well as the diagonal Ising interaction for two spin-1/2 particles [\citenum{reif2009}, \S 10.6]. We emphasize that the parameters $\omega^{(j)},\coupstr$ are all time-independent as we focus on the autonomous paradigm. For the universe Hilbert space, we adopt the induced basis with usual binary labels, $\left\lbrace \ket{n}\right\rbrace_{n=0,\hdots,3}$, such that $\ket{0} := \ket{00} := \ket{0^A}\ket{0^B},\ket{1} := \ket{01}$ etc. [\citenum{nielsen2010}, \S 4.6]. At $t=0$, the universe state is a tensor product and the reduced states must be fully general in order to address the dynamical maps for both parties (see \ref{subsec:min_dissip_approach}). We denote them	
		
		\begin{equation}
		\label{eq:canon_deph:local_states_init}
			\varrho^{(j)}(0) = 
			\begin{pmatrix}
				p_0^{(j)} & \varrho_{01}^{(j)}(0) \\
				* & p_1^{(j)}
			\end{pmatrix}
		\end{equation}
	where the implicit term $(*)$ is fulfilled by hermiticity, and of course $p_0^{(j)}+p_1^{(j)}=1$. Note that $p_k^{(j)}:=\varrho_{kk}^{(j)}(0)\in\mathbb{R}$, for clarity of notation, in benefit from the diagonal terms being constants under the Hamiltonian \eqref{eq:canon_deph:hint_deph}. Indeed, in Appendix \ref{sec:appx:solving_dynamics} we solve the Liouville-von Neumann equation for the universe dynamics and, by partial-tracing the global state $\rho(t)$, we show that the evolved reduced states are
   
		\begin{equation}
		\label{eq:canon_deph:local_states_sol}
			\varrho^{(j)}(t) = 
			\begin{pmatrix}
				p_0^{(j)} & 
				e^{
					i\omega^{(j)} t
				}g^{(j)}(t)
				\varrho_{01}^{(j)}(0) \\
				* & p_1^{(j)}
			\end{pmatrix},
		\end{equation}
	where
		
		\begin{equation}
		\label{eq:canon_deph:def_g_ab}
		\begin{aligned}
            g^A(t) &:=  
			p^B_0
			e^{-2i\coupstr t}+
			p^B_1
			e^{2i\coupstr t},\\
			g^B(t)&:=
			p^A_0e^{-2i\coupstr t}+
			p^A_1
			e^{2i\coupstr t}.
		\end{aligned}
		\end{equation}
    
    As expected, $\varrho_{kk}^{(j)}(t)=\varrho_{kk}^{(j)}(0)=:p_k^{(j)}$. The natural-frequency rotation of the coherence $\varrho^{(j)}_{01}(t)$ on the $\mathbb{C}$-plane is modulated by $g^{(j)}(t)$, which oscillates with frequency $2\coupstr$ and contains the influence of the ``environment'' initial state (recall equation \ref{eq:canon_deph:def_g_ab}). Moreover, as the interaction $H^I$ \eqref{eq:canon_deph:hint_deph} is such that $\left[H, H^{(j)}\right]=\left[H, H^I\right]=0$, each piece in the universe Hamiltonian \eqref{eq:canon_deph:h_tot_def} has constant average value.\footnote{
    	In particular, this model satisfies strict energy conservation: $\avg{H^A}+\avg{H^B} = \text{constant}$, and some authors would address thermodynamics based on these quantities. Nevertheless, here we take advantage of this simple model to probe the far more general scheme of \cite{Hayden_Sorce_2022,Colla_Breuer_2022}, which, as we will soon show, entails non-trivial corrections for the local energies even in this case.
    } From $H^{(j)} =
    \left( \omega^{(j)}/2 \right) \sigma_z^{(j)}$ we find
    
    	\begin{equation}
    	\label{eq:canon_deph:avg_h_j}
    		\avg{H^{(j)}} = 
    		\frac{\omega^{(j)}}{2} 
    		\left(p^{(j)}_1-p^{(j)}_0\right).
    	\end{equation}
    
    It is particularly simple to calculate $\avg{H^I}$ at $t=0$:
    $
    \tr{\left(H^I\rho(0)\right)}=
    \tr{\left(
    	\coupstr \sigma_z^A\otimes\sigma_z^B
    	\varrho^A(0)\otimes\varrho^B(0)
    	\right)}
    =\coupstr \avg{\sigma_z^A}\avg{\sigma_z^B}
    $, whence
    
    	\begin{equation}
    	\label{eq:canon_deph:avg_h_int}
    		\avg{H^I}=
    		\coupstr \left(
    		p^{A}_1-p^{A}_0
    		\right)
    		\left(
    		p^{B}_1-p^{B}_0
    		\right).
    	\end{equation}
    
    One may double-check this result with $\rho(t)$ \eqref{eq:solving_dynamics:full_state}.
    
\subsection{Exact master equations}

	Having obtained the exact form of the reduced states along time, we can unambiguously determine the open-system dynamical generator $\mathcal{L}^{(j)}_t$ (equation \ref{eq:canon_deph:master_eqn_abstract}) and cast it in the unique form described above (sec. \ref{subsec:min_dissip_approach}). On Appendix \ref{sec:appx:obtain_eqns} we derive the equations
	
		\begin{equation}
		\begin{aligned}
			\label{eq:canon_deph:exact_master_equation}
			\dot{\varrho}^{(j)}(t) &=	
			\mathcal{L}^{(j)}_t
			\left\lbrace 
			\varrho^{(j)}(t)
			\right\rbrace	
			\\&= 
			-i\left[ 
			\tilde{H}^{(j)}(t), \varrho^{(j)}(t) 
			\right]
			+ \mathcal{D}^{(j)}_t\left[
			\varrho^{(j)}(t) 
			\right],
		\end{aligned}
		\end{equation}
	where 
	
		\begin{align}
			\tilde{H}^{(j)}(t) &=
			\frac{\tilde{\omega}^{(j)}(t)}{2}\sigma_z^{(j)},
			\label{eq:canon_deph:effective_ham} \\
			\tilde{\omega}^{(j)}(t) &=
			\omega^{(j)} + 
			\Im{\left(
				\frac{
					\dot{g}^{(j)}(t)	
				}{
					g^{(j)}(t)
				}
			\right)}  
			\label{eq:canon_deph:effective_omega},
		\end{align}	
	and
		\begin{align}
			\mathcal{D}^{(j)}_t\left[\cdot \right]
			&=
			\gamma^{(j)}(t)
			\left( 
			\sigma^{(j)}_z \cdot  \sigma^{(j)\dagger}_z - \frac{1}{2}\left\lbrace \sigma^{(j)\dagger}_z\sigma_z^{(j)}, \cdot  \right\rbrace
			\right),
			\label{eq:canon_deph:dissip}\\
			\gamma^{(j)}(t)&=
			-\frac{1}{2}
			\Re{\left(
				\frac{\dot{g}^{(j)}(t)}{g^{(j)}(t)}
				\right)}
			\label{eq:canon_deph:effective_gamma}.
		\end{align}
	
	Here, $\Re$ and $\Im$ denote real and imaginary parts, respectively. Equation \eqref{eq:canon_deph:exact_master_equation} displays the map $\mathcal{L}^{(j)}_t$ in the standard form \eqref{eq:canon_deph:gen_master_eq_decomp}, with a single-term dissipative part \eqref{eq:canon_deph:dissip} already in Lindblad form  \eqref{eq:canon_deph:gen_dissip_lindb} with traceless jump operator $L_k^{(j)}=\sigma_z^{(j)}$. The unitary-dissipative decomposition addressed must be the unique one with these properties, as shown in \cite{Hayden_Sorce_2022} (recall \ref{subsec:min_dissip_approach}). We conclude that the traceless Hermitian operator \eqref{eq:canon_deph:effective_ham} with effective frequency \eqref{eq:canon_deph:effective_omega} is the canonical Hamiltonian of the minimal dissipation approach for the model described here (\ref{subsec:model_two_qubit_deph}). 
	
	We have then explicitly determined the exact master equations for the two parties within a closed, bipartite quantum system. A crucial requirement for that is to allow an arbitrary initial state of the ``environment'' (here, system $B$), so that the map $\Phi^B_t$ is completely specified and the procedure described in \ref{subsec:min_dissip_approach} is well-defined for $B$ as well as for $A$.  The choice of a finite-dimensional system $B$ (indeed, $\dim{\mathcal{H}^B} = \dim{\mathcal{H}^A} = 2$) and the simple coupling model \eqref{eq:canon_deph:hint_deph} counterbalanced the generality of the initial state, allowing us to make all the way through the Lindblad-form master equation \eqref{eq:canon_deph:exact_master_equation} -- besides, with analytic expressions for the coefficients. Indeed, recalling the definitions \eqref{eq:canon_deph:def_g_ab}, one may reduce \eqref{eq:canon_deph:effective_omega} to 
	
		\begin{equation}
			\label{eq:canon_deph:omega_eff_explicit_j}
			\tilde{\omega}^{(j)}(t) =
			\omega^{(j)} + 
			\frac{
				2\coupstr \left(
					p^{(\sim j)}_1-p^{(\sim j)}_0
				\right)
			}{
				1-
				4p^{(\sim j)}_0p^{(\sim j)}_1
				\sin ^2\left( 2\coupstr t \right)
			}
		\end{equation}
	where ${\sim} j$ denotes the complementary part of $j$ (${\sim} A = B, {\sim} B = A$).

\subsection{Thermodynamics on the minimal dissipation approach}

     We can now evaluate the internal energies $U^{(j)}(t) = \avg{\tilde{H}^{(j)}(t)} = \tr{\left(\varrho^{(j)}(t)\tilde{H}^{(j)}(t)\right)}$. 

     Equations \eqref{eq:canon_deph:local_states_sol} and \eqref{eq:canon_deph:effective_ham} yield $U^{(j)}(t) =
     \tilde{\omega}^{(j)}(t)\left( p^{(j)}_1-p^{(j)}_0\right)/2$, analogous to \eqref{eq:canon_deph:avg_h_j}. Now by substituting equation \eqref{eq:canon_deph:omega_eff_explicit_j} we can write
	
		\begin{equation}
			\label{eq:canon_deph:u_j_explicit}
			U^{(j)}(t) =
			\avg{H^{(j)}} + 
			\frac{
				\coupstr 
				\left(p^{(j)}_1-p^{(j)}_0\right)
				\left(
				p^{(\sim j)}_1-p^{(\sim j)}_0
				\right)
			}{
				1-
				4p^{(\sim j)}_0p^{(\sim j)}_1
				\sin ^2\left( 2\coupstr t \right)
			},
		\end{equation}
	which, by \eqref{eq:canon_deph:avg_h_int}, may be simplified to
	
	\begin{equation}
		\label{eq:canon_deph:u_j_explicit_2}
		U^{(j)}(t) =
		\avg{H^{(j)}} + 
		\frac{
			\avg{H^I}
		}{
			1-
			4p^{(\sim j)}_0p^{(\sim j)}_1
			\sin ^2\left( 2\coupstr t \right)
		},
	\end{equation}
	for each subsystem $j$.

\begin{figure}
	\centering
	\vspace{-35pt}
	\includegraphics[width=\linewidth]{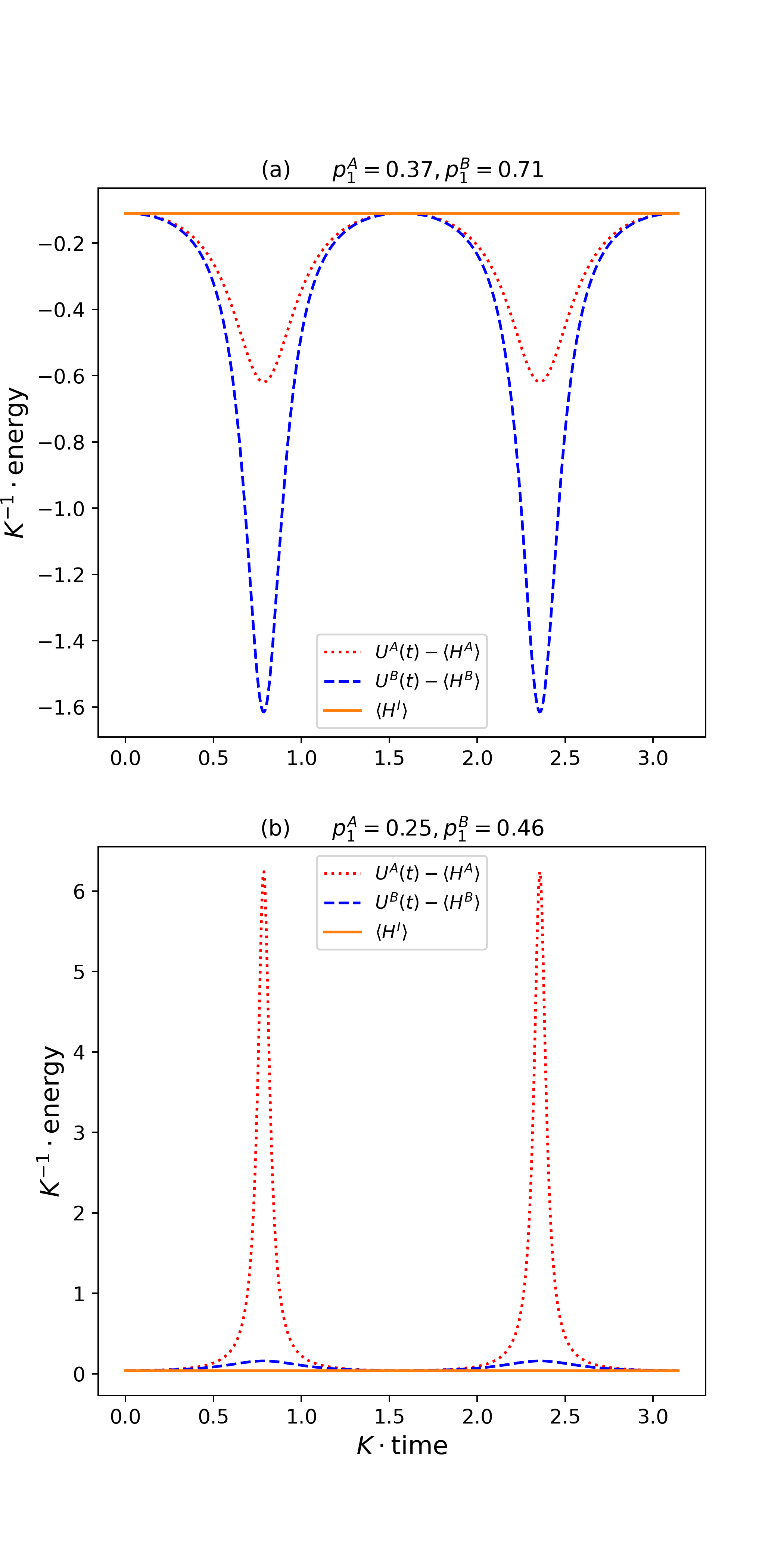}\vspace{-35pt}
	\caption{Time-dependent internal energy corrections for the two parts of the closed universe described in \ref{subsec:model_two_qubit_deph}, according to the minimal dissipation approach \cite{Hayden_Sorce_2022,Colla_Breuer_2022}, for two particular choices of the initial populations $p_1^A,p_1^B$. See equation \eqref{eq:canon_deph:u_j_explicit_2}. One can see that, in this model, both internal energies oscillate ``in phase'', rather than counter-balancing each other. Moreover, their amplitudes are generally distinct (equation \ref{eq:canon_deph:u_ab_peaks}) and, in particular, figure (b) illustrates how the peaks in $U^A(t)$ become arbitrarily large if $p_1^B$ is taken sufficiently close to $1/2$.}
	\label{fig:u_ab_illust}
\end{figure}

	Some observations are now possible. The dynamically defined open-system internal energy $U^{(j)}$, derived from the ``canonical'' effective Hamiltonian (Section \ref{subsec:min_dissip_approach}), for this particular model, turns out to be precisely the averaged ``bare'' Hamiltonian $H^{(j)}$ plus a coupling-strength-proportional term, in which we could even identify a factor $\avg{H^I}$. Now, as the equations all hold for both $j$, we can study whether a change in $U^A(t)$ corresponds to an opposite change in $U^B$(t) -- namely, whether some form of energy additivity holds in the minimal dissipation approach. It turns out, as we can read from \eqref{eq:canon_deph:u_j_explicit_2}, that both $U^A(t)$ and $U^B(t)$ oscillate ``in phase'', while only the multiplying factor of the sinusoid $\sin^2\left(2\coupstr t\right)$ depends on the chosen system. See figure \ref{fig:u_ab_illust}(a). Indeed, the sum of internal energies is
	
		\begin{widetext}
		\begin{equation}
			U^A(t) + U^B(t) = 
			\avg{H^L} + 
			\avg{H^I}\left(
			\frac{
				1
			}{1 -4p^A_0p^A_1\sin ^2\left( 2\coupstr t \right)}
			+
			\frac{
				1
			}{1 -4p^B_0p^B_1\sin ^2\left( 2\coupstr t \right)}
			\right),
		\end{equation}
		\end{widetext}
	which besides being different from $\avg{H}$ is clearly time-dependent. In other words, neither of Definitions \ref{def:weak_add} and \ref{def:str_add_select} is satisfied by the minimal-dissipation EHR. Curiously, even at $t=0$ the correction over $\avg{H^L}$ is \textit{twice} $\avg{H^I}$. Both systems gain or lose internal energy at the same time intervals, although their joint energy is a well-known constant.
	
	Our focus here is internal energy, the fundamental quantity for energy exchange. However, in this example it is straightforward to extend the analysis to work and heat as defined in \cite{Colla_Breuer_2022}. Equations \eqref{eq:canon_deph:local_states_sol} and \eqref{eq:canon_deph:effective_ham} imply $\dot{Q}^{(j)}(t):= \tr{\left[\dot{\varrho}^{(j)}(t)\tilde{H}^{(j)}(t)\right]}=0$ -- namely, no heat as the populations are constants; then for each subsystem $\Delta W^{(j)}(t) = U^{(j)}(t) - U^{(j)}(0)$, and we are forced to conclude that at each half-period $A$ and $B$ either perform work on each other simultaneously, or receive work from each other simultaneously.
	
	We may also study the values of the peaks in $U^{(j)}(t)$ (dismissing the sign). They happen when $\sin^2\left(2\coupstr t\right)=1$, so from \eqref{eq:canon_deph:u_j_explicit} we obtain	
	
		\begin{equation}
			\label{eq:canon_deph:u_ab_peaks_passage}
			\max_t \left| 
				\frac{
					U^{(j)}(t) - \avg{H^{(j)}}
				}{\coupstr}
			\right|  =
			\frac{
				\left|
					p^{(j)}_1-p^{(j)}_0
				\right|
				\left|
					p^{(\sim j)}_1-p^{(\sim j)}_0
				\right|
			}{
				1-
				4p^{(\sim j)}_0p^{(\sim j)}_1
			}.
		\end{equation}
	Recalling that $p_0^{(j)}=1-p_1^{(j)}$, the expression above may be recast as
	
		\begin{equation}
			\label{eq:canon_deph:u_ab_peaks}
			\max_t \left| 
			\frac{
				U^{(j)}(t) - \avg{H^{(j)}}
			}{\coupstr}
			\right|  =
			\left|
			\frac{
				p^{(j)}_1-1/2
			}{
				p^{(\sim j)}_1-1/2
			}
			\right|,
		\end{equation}
	insofar as $p_1^{(\sim j)}\neq1/2$. The peaks are thus unbounded: if $p_1^{(\sim j)}$ is taken close enough to $1/2$, the peak value in $U^{(j)}(t)$ becomes arbitrarily large. We stress that the initial states of $A$ and $B$ are independent and thus this procedure may be done for a fixed $p_1^{(j)}$. This singular behavior may be imputed to the dynamical map $\Phi^{(j)}_t$ failing to be invertible at the peak times if (and only if) $p_1^{(\sim j)}=1/2$, as per Appendix \ref{subsec:appx_obtain_eqns_exist}. However, we stress that, for \textit{any} $\varepsilon:= p_1^{(\sim j)}-1/2\neq 0$, no matter how small, the map $\Phi^{(j)}_t$ is invertible for all times and therefore the master equation \eqref{eq:canon_deph:exact_master_equation} exists and the approach of \cite{Hayden_Sorce_2022, Colla_Breuer_2022} should apply. Thus, for instance, if $p^B_1-1/2$ is close to (but different from) $0$, the effective internal energy $U^A(t)$, though always well-defined, becomes arbitrarily large at peak times, which is not observed for $U^B(t)$ provided that $p^A_1$ is does not follow the behavior of $p^B_1$ (namely, if $\left|p^A_1-1/2\right|$ is moderately large). See Figure \ref{fig:u_ab_illust}(b) as an illustration.
	
	In this Section, we have formally characterized the minimal-dissipation quantum thermodynamical approach of ref. \cite{Colla_Breuer_2022}, and then analyzed its predictions for the particular case of a two-qubit closed quantum universe, in which the parts are coupled through a population-preserving (commuting) Hamiltonian \eqref{eq:canon_deph:hint_deph}. We have seen that none of the notions of additivity introduced in Sec. \ref{sec:defs_addit_quant} is satisfied by this EHR-- not even the weak form, which is just a statement of energy conservation. In particular, the minimal-dissipation approach predicts that each subsystem simultaneously acts as a work source (or sink) for the other, although no third party exists in the model. Moreover, the mismatch between the sum of effective internal energies $U^{A}(t)+U^B(t)$ and the known universe energy $\avg{H}$ becomes arbitrarily large if either of the initial populations gets close to the singular point $p^{(j)}_k=1/2$. Hopefully, these results illustrate the consequences of addressing definitions of internal energy that are not additive either by definition or by construction.  
	


\section{Conclusions}
\label{sec:conclusions}

	Autonomous-universe quantum thermodynamics deals with energy exchange between genuine open quantum systems, with no room for a classical external agent. Along with strong coupling and correlations, this paradigm may be considered to draw the currently active border in the scope of quantum thermodynamics, moving ever further away from the traditional semi-classical paradigm of a quantum system surrounded by a large heat bath and subject to a classical source of work \cite{Alicki_1979}. 
	
	In this realm, several definitions of work and heat have been proposed \cite{Weimer_etal_2008, Alipour_etal_2016, Colla_Breuer_2022, Pyharanta_etal_2022, Valente_etal_2018, Malavazi_Brito_2022, Huang_Zhang_2022, Ali_etal_2020, Alipour_etal_2022, Ahmadi_etal_2023, Silva_Angelo_2023}, which may overshadow an underlying disagreement on how to generally calculate the \textit{internal energy} of an open system. Such fundamental quantity is most commonly defined as the quantum mechanical average of either the system's bare Hamiltonian \cite{Ali_etal_2020, Alipour_etal_2022, Ahmadi_etal_2023} or an assigned effective Hamiltonian \cite{Weimer_etal_2008, Alipour_etal_2016, Colla_Breuer_2022, Pyharanta_etal_2022, Valente_etal_2018, Malavazi_Brito_2022, Huang_Zhang_2022}. The latter approach may be mapped into the concept that an open system embodies part of the interaction energy between itself and its surroundings \cite{Alipour_etal_2016, Dou_etal_2018, Huang_Zhang_2022, Seegebrecht_Schilling_2024}, in agreement with the usual interpretation of the Lamb shift \cite{Welton_1948}. However, with rare exceptions \cite{Alipour_etal_2016,Malavazi_Brito_2022}, most specific frameworks in this trend address effective Hamiltonians explicitly only for the system of interest, without an account of the corresponding energy corrections for the environment.  It cannot be overemphasized that, within the autonomous universe paradigm, both ``system'' and ``environment'' (or subsystems $A$ and $B$) comprise a closed system (universe), whose internal energy is given by the average value of a well-defined observable, namely the total Hamiltonian.
	
	In this article, we have sustained that, in order to allow physically reasonable definitions of ``renormalized'' open-system energies, one must apply the same definitions for the two parts of the closed universe and check for additivity. 
	We started by addressing an abstract framework that essentially fits any approach to defining effective Hamiltonians for the two parts of a closed autonomous universe (Section \ref{sec:defs_addit_quant}), the key notion being that of an effective Hamiltonian rule (EHR), Def. \ref{def:eff_ham_rule}. With this tool, we rigorously defined additivity, in both a weak and a strong version, the latter containing some subtleties regarding the ground-state energy of the universe Hamiltonian, which was discussed. Connections with preceding works were presented, and the Schmidt-decomposition approach of reference \cite{Malavazi_Brito_2022}, suitable for pure universe states, was identified as strongly additive by invariance.
	As an application, in Section \ref{sec:canon_deph}, we studied the minimal-dissipation EHR of references \cite{Hayden_Sorce_2022,Colla_Breuer_2022}. After noting that its very structure is incompatible with strong additivity in the invariant form (Proposition \ref{prop:nogo_dynam_reduc}), we further investigated it by means of a simple model of two qubits coupled via a commuting interaction. Having obtained the exact master equations for both open systems,  we found that this particular EHR is not additive even in the weak sense (a mere conservation of the total energy), and obtained some expressions to illustrate the consequences of that. Specifically, in the studied case, both $A$ and $B$ deliver energy to (or receive energy from) each other at the same time, and further, all of the energy exchange in this case is labeled as work. Moreover, the mismatch between the total effective internal energy and the well-known closed-system energy of the universe can become arbitrarily large, depending on the initial conditions. 

	Various quantum thermodynamical frameworks have been proposed in the autonomous universe paradigm. Although many of them agree in adopting some effective Hamiltonian, the several methods for defining this operator certainly entail conflicting quantifications of internal energy, and only very recently authors started to systematically compare these definitions \cite{Dann_Kosloff_2023,Seegebrecht_Schilling_2024}. Our goal here was not to address one more competing definition, nor to endorse any of the available; rather, we intend to raise the attention to an important physical property that has been neglected in almost all of them. We propose additivity as a basic consistency check, since in the autonomous paradigm the energy of the compound system (``universe'') is essentially well-known. The endeavor of devising a completely general, scale-independent theoretical framework comprising thermodynamics and quantum physics in full glory should either succeed or find a clear border in its range of validity. If the former is to be the case, fundamental physical principles such as energy conservation should lie at the heart of the theory.

\section*{Acknowledgments}

	L.R.N.'s work was partially supported by the São Paulo Research Foundation (FAPESP), Grant No. 2021/01365-9. F. B. acknowledges partial support from Brazilian National Institute of Science and Technology of Quantum Information (CNPq INCT-IQ 465469/2014-0) and from Conselho Nacional de Desenvolvimento Cient\'ifico e Tecnol\'ogico (402074/2023-8). The authors also thank Dr. André H. Malavazi for critical reading of the manuscript.

\appendix

\section{Decomposition of a bipartite-universe Hamiltonian}
\label{sec:app:decomp_H}

In the same framework introduced in \ref{sec:defs_addit_quant}, let the Hilbert spaces $H^A, H^B$ have respective dimensions $d^A$ and $d^B$ (possibly infinite). Then $\text{dim}\left(\mathcal{H}\right)=d^A d^B$. Also, $\mathcal{B}\left(\mathcal{H}\right)$ is a real vector space of dimension $\left(d^Ad^B\right)^2$. As usual (see \cite{Hioe1981,schlienz1995,Alipour_etal_2020}, and [\citenum{Breuer_Petruccione_2002}, \S 2.2.1], for example), we adopt a basis
$
	\left\lbrace
	A_k;\hspace{5pt}
	k=0,...,(d^A)^2-1 
	\right\rbrace
$
for $\mathcal{B}\left(\mathcal{H}^A\right)$, such that $A_0=\mathbb{I}^A$ and  $\tr\left(A_{\ell}A_k\right)\propto \delta_{\ell k}$ (orthogonality). It follows that $\tr{A_k}=0, \forall k \neq 0$. We adopt a basis $\set{B_k}$ for $\mathcal{B}\left(\mathcal{H}^B\right)$ with the same properties. The two local bases induce one for $\mathcal{B}\left(\mathcal{H}\right)$, namely
$
	\left\lbrace 
	A_\ell \otimes B_k;\hspace{5pt}
	\ell=0,...,(d^A)^2-1;
	k=0,...,(d^B)^2-1
	\right\rbrace
$.

We then expand any Hamiltonian $H\in\mathcal{B}\left( \mathcal{H} \right)$ as 

\begin{equation}
	H = 
	\sum_{\ell=0}^{(d^A)^2-1}
	\sum_{k=0}^{(d^B)^2-1} 
	h_{\ell k}A_\ell\otimes B_k 
\end{equation}
which we can recast as 

\begin{equation}
	\begin{gathered}
		H=
		h_{00}\mathbb{I} + 
		\sum_{\ell=1}^{(d^A)^2-1} h_{\ell 0} 
		A_\ell \otimes\mathbb{I}^B +
		\sum_{k=1}^{(d^B)^2-1} h_{0k}
		\mathbb{I}^A
		\otimes B_k \\
		+\sum_{\ell=1}^{(d^A)^2-1}
		\sum_{k=1}^{(d^B)^2-1} 
		h_{\ell k}A_\ell\otimes B_k .  
	\end{gathered}
\end{equation}

Note that, although the adopted bases  $\left\lbrace A_k\right\rbrace,\left\lbrace B_k\right\rbrace$ are arbitrary up to the postulated properties, the resulting decomposition of $H$ in four parts as above is not, as it relies merely on the fixed points $A_0 = \mathbb{I}^A,B_0=\mathbb{I}^B$. In fact, due to the trace properties imposed for the local bases, we can write, in the finite-dimensional case,

\begin{align}
	h_{00} &= \frac{1}{d^Ad^B}
	\tr H, \\
	\sum_{\ell=1}^{(d^A)^2-1}h_{\ell0}A_\ell 
	&= \frac{1}{d^B}\tr_B H - h_{00}\mathbb{I}^A, \\
	\sum_{k=1}^{(d^B)^2-1}h_{0k}B_k 
	&= \frac{1}{d^A}\tr_A H - h_{00}\mathbb{I}^B,
\end{align}
and for each equation above the right-hand side is explicitly basis-independent. We can then define the local and interaction parts, 

\begin{gather}
	H^L = h_{00}\mathbb{I} + 
	\sum_{\ell=1}^{(d^A)^2-1} h_{\ell 0} 
	A_\ell\otimes\mathbb{I}^B +
	\sum_{k=1}^{(d^B)^2-1} h_{0k}
	\mathbb{I}^A
	\otimes B_k,\\
	H^I = \sum_{\ell=1}^{(d^A)^2-1}
	\sum_{k=1}^{(d^B)^2-1} 
	h_{\ell k}A_\ell\otimes B_k,
\end{gather} 
and the decomposition $H = H^L + H^I$ is clearly general and unique. 

In turn, the matter of defining the bare Hamiltonian of each part, i. e. finding the decomposition $H^L = H^A\otimes\mathbb{I}^B + \mathbb{I}^A\otimes H^B$, is a little more subtle in the general case, $h_{00}\neq 0$, as an additional criterion must be introduced to split the identity term $h_{00}\mathbb{I}=h_{00}\mathbb{I}^A\otimes\mathbb{I}^B$ in two. On the other hand, if such an identity term was fixed by either of the approaches (\ref{item:selective_traceless}, \ref{item:selective_groundzero}) described in \ref{sec:defs_addit_quant}, then the ambiguity disappears: in the traceless convention, one has $h_{00}=0$; otherwise, in the zero-ground-energy convention, a particular identity term is needed to set the ground energy of each $H^{(j)}$ to zero.

\section{Solving the dynamical equations (Section \ref{sec:canon_deph})}
\label{sec:appx:solving_dynamics}

	In matrix form on the basis introduced in sec. \ref{subsec:model_two_qubit_deph}, the total Hamiltonian is 
		
		\begin{equation}
			H = 
			\text{diag}\left( 
				\coupstr -\omega,
				-\coupstr +\delta,
				-\coupstr-\delta,
				\coupstr+\omega 
			\right),
		\end{equation}
	where for clarity we introduce the notation $\omega := \left( \omega^A+\omega^B \right)/2$,
	$\delta :=\left( \omega^B-\omega^A \right)/2$. We emphasize that in our representation
	
		\begin{equation}
		\label{eq:solving_dynamics:pauli}
		\begin{gathered}
			\sigma_x = 
			\begin{pmatrix}
				0 & 1 \\
				1 & 0
			\end{pmatrix},
			\sigma_y =
			\begin{pmatrix}
				0 & i \\
				-i & 0
			\end{pmatrix},
			\sigma_z =
			\begin{pmatrix}
				-1 & 0 \\
				0 & 1
			\end{pmatrix},
		\end{gathered}
		\end{equation}	
	so that the ground state ($k=0$) comes first and has lowest eigenvalue in $H^{(j)}\propto \sigma_z^{(j)}$. For a generic universe density matrix $\rho = \sum_{m,n=0}^3\rho_{mn}\ket{m}\bra{n}$, the Liouville-von Neumann equation $\dot{\rho}=-i\left[H,\rho\right]$ then takes the form
	
		\begin{widetext}
		\begin{equation}
		\begin{pmatrix}
			\dot{\rho}_{00} & 
			\dot{\rho}_{01} & 
			\dot{\rho}_{02} & 
			\dot{\rho}_{03} \\
			\dot{\rho}_{10} & 
			\dot{\rho}_{11} & 
			\dot{\rho}_{12} & 
			\dot{\rho}_{13} \\
			\dot{\rho}_{20} & 
			\dot{\rho}_{21} & 
			\dot{\rho}_{22} & 
			\dot{\rho}_{23} \\
			\dot{\rho}_{30} & 
			\dot{\rho}_{31} & 
			\dot{\rho}_{32} & 
			\dot{\rho}_{33}
		\end{pmatrix}=
		-i
		\begin{pmatrix}
			0 & 
			\left(
			2\coupstr -\omega^B
			\right)\rho_{01}&
			\left(
			2\coupstr - \omega^A 
			\right)\rho_{02}&
			-2\omega\rho_{03} \\
			\left(
			-2\coupstr +\omega^B
			\right)\rho_{10}&
			0&
			2\delta\rho_{12} &
			-\left(
			2\coupstr+\omega^A
			\right)\rho_{13}\\
			\left(
			-2\coupstr +\omega^A 
			\right)\rho_{20}&
			-2\delta\rho_{21}&
			0&
			-\left(
			2\coupstr+\omega^B
			\right)\rho_{23}\\
			2\omega\rho_{30} &
			\left(
			2\coupstr+\omega^A
			\right)\rho_{31}&
			\left( 
			2\coupstr+\omega^B
			\right)\rho_{32}&
			0
		\end{pmatrix}
		\end{equation}
	with straightforward solution
	
		\begin{equation}
		\label{eq:solving_dynamics:full_state}
		\rho(t) = 
		\begin{pmatrix}
			\rho_{00}(0) & 
			\rho_{01}(0)
			e^{-i\left(
				2\coupstr -\omega^B
				\right)t} & 
			\rho_{02}(0)
			e^{-i\left(
				2\coupstr - \omega^A 
				\right)t} & 
			\rho_{03}(0)e^{+2i\omega t} \\
			* & 
			\rho_{11}(0) & 
			\rho_{12}(0)e^{-2i\delta t} & 
			\rho_{13}(0)e^{i\left(
				2\coupstr+\omega^A
				\right)t} \\
			* & 
			* & 
			\rho_{22}(0) & 
			\rho_{23}(0)e^{i\left(
				2\coupstr+\omega^B
				\right)t} \\
			* & 
			* & 
			* & 
			\rho_{33}(0) 
		\end{pmatrix},
		\end{equation}
		\end{widetext}
	in terms of a still arbitrary initial state. By taking the partial traces, we obtain for the reduced density matrices: $\varrho^A_{00}(t) = \rho_{00}(0)+\rho_{11}(0)$;
	$\varrho^A_{11}(t) =	\rho_{22}(0)+\rho_{33}(0)$; 
	$\varrho^B_{00}(t) =\rho_{00}(0)+\rho_{22}(0)$;
	$\varrho^B_{11}(t) =\rho_{11}(0)+\rho_{33}(0)$; and 

		\begin{equation}
		\begin{aligned}
		\label{eq:solving_dynamics:coherences_1}
			\varrho^A_{01}(t) &=
			e^{i\omega^A t}
			\left(
			\rho_{02}(0)e^{-2i\coupstr t}+
			\rho_{13}(0)e^{2i\coupstr t}
			\right), \\
			\varrho^B_{01}(t) &=
			e^{i\omega^B t}
			\left(
			\rho_{01}(0)e^{-2i\coupstr t}+
			\rho_{23}(0)e^{2i\coupstr t}
			\right). 
		\end{aligned}
		\end{equation}
	
	As per sec. \ref{subsec:min_dissip_approach}, we are interested in the particular case of a product initial state, $\rho(0)=\varrho^A(0)\otimes\varrho^B(0)$. Then each $\rho_{ij}(0)$ is written in terms of reduced density matrix elements, which is achieved by assigning to each index $i=0,...,3$ the correspondent binary representation, and then applying the tensor product relation  $\rho_{ij;kl}(0)=\varrho^A_{ik}(0)\varrho^B_{jl}(0)$. Crucially, we have
	
	\begin{equation}
	\begin{aligned}
		\rho_{02}(0) &= \rho_{00;10}(0) = \varrho^A_{01}(0)\varrho^B_{00}(0)\\
		\rho_{13}(0) &= 
		\rho_{01;11}(0) = 
		\varrho^A_{01}(0)
		\varrho^B_{11}(0)\\
		\rho_{01}(0) &= 
		\rho_{00;01}(0) = 
		\varrho^A_{00}(0)
		\varrho^B_{01}(0)\\
		\rho_{23}(0) &= 
		\rho_{10;11}(0) = 
		\varrho^A_{11}(0)
		\varrho^B_{01}(0)
	\end{aligned}
	\end{equation}
	and equation \eqref{eq:solving_dynamics:coherences_1} becomes
	
		\begin{equation}	
			\begin{aligned}
				\label{eq:solving_dynamics:coherences_2}
				\varrho^A_{01}(t) &=
				e^{i\omega^A t}
				\left(
				\varrho^B_{00}(0)
				e^{-2i\coupstr t}+
				\varrho^B_{11}(0)
				e^{+2i\coupstr t}
				\right)
				\varrho^A_{01}(0), \\
				\varrho^B_{01}(t) &=
				e^{i\omega^B t}
				\left(
				\varrho^A_{00}(0)
				e^{-2i\coupstr t}+
				\varrho^A_{11}(0)
				e^{+2i\coupstr t}
				\right)
				\varrho^B_{01}(0). 
			\end{aligned}
		\end{equation}

	With the additional notations for the constant diagonal terms, $p_k^{(j)}:=\varrho_{kk}^{(j)}(0)=\varrho_{kk}^{(j)}(t)$, and the functions $g^{(k)}(t)$ defined in \eqref{eq:canon_deph:def_g_ab}, we finally obtain the forms \eqref{eq:canon_deph:local_states_sol} for the local states.
	
\section{Obtaining the exact master equations (Section \ref{sec:canon_deph})}
\label{sec:appx:obtain_eqns}

\subsection{Deriving the equations}
	
	In order to obtain the exact master equations underlying the local states \eqref{eq:canon_deph:local_states_sol}, the standard approach involves explicit differentiation and inversion of the local dynamical maps $\Phi^{(j)}_t$ (see e.g. \cite{Smirne_Vacchini_2010}). Here we take an alternative course, though equivalent. As equation \eqref{eq:canon_deph:local_states_sol} contains the action of the dynamical map on the general initial state \eqref{eq:canon_deph:local_states_init}, it completely specifies the map -- as represented on the specific basis  $\left\lbrace \ket{n}\right\rbrace_{n=0,\hdots,3}$ of $\mathcal{H}$. Indeed we can write 

		\begin{equation}
		\label{eq:obtain_eqns:rho_j}
			\Phi^{(j)}_t\left\lbrace 
			\varrho^{(j)}(0)
			\right\rbrace=
			\varrho^{(j)}(t)=
			\begin{pmatrix}
			p_0^{(j)} &
			f^{(j)}(t)\varrho^{(j)}_{01}(0)\\
			*&
			p_1^{(j)}
			\end{pmatrix}
		\end{equation}
	where, for compactness, 
	$f^{(j)}(t):=e^{ i\omega^{(j)} t }g^{(j)}(t)$. Taking the time derivative,
		
		\begin{equation}
		\label{eq:obtain_eqns:dot_rho_j}
			\dot{\Phi}^{(j)}_t\left\lbrace 
			\varrho^{(j)}(0)
			\right\rbrace
			= 
			\dot{\varrho}^{(j)}(t) =
			\begin{pmatrix}
			0 &
			\dot{f}^{(j)}(t) \varrho^{(j)}_{01}(0)\\
			* & 0 
			\end{pmatrix}
		\end{equation}

	Next step is to replace $\varrho^{(j)}(0)$ with $\varrho^{(j)}(t)$, which corresponds to inverting $\Phi^{(j)}_t$. Here, due to the simplicity of \eqref{eq:obtain_eqns:rho_j} we can just write $\varrho^{(j)}_{01}(t) = f^{(j)}(t)\varrho^{(j)}_{01}(0)$, whence 

		\begin{equation}
		\label{eq:obtain_eqns:invert_01}
			\varrho^{(j)}_{01}(0) = \frac{1}{f^{(j)}(t)}
			\varrho^{(j)}_{01}(t).
		\end{equation}

	At this point we just suppose $f^{(j)}(t)\neq 0$, a hypothesis discussed on Section \ref{subsec:appx_obtain_eqns_exist}. Substituting the above in equation \eqref{eq:obtain_eqns:dot_rho_j}, we have 

		\begin{equation}
		\label{eq:obtain_eqns:dot_rho_j_inv}
		\begin{gathered}
			\dot{\Phi}^{(j)}_t
			\circ\left(
			\Phi^{(j)}_t
			\right)^{-1}
			\left\lbrace 
			\varrho^{(j)}(t)
			\right\rbrace 
			=
			\dot{\varrho}^{(j)}(t) \\
			= 
			\begin{pmatrix}
				0 &
				\varphi^{(j)}(t) \varrho^{(j)}_{01}(t)\\
				\varphi^{(j)*}(t) \varrho^{(j)}_{10}(t) & 0 
			\end{pmatrix}
		\end{gathered}
		\end{equation}
	where
		\begin{equation}
		\label{eq:obtain_eqns:def_varphi}
			\varphi^{(j)}(t):=
			\frac{\dot{f}^{(j)}(t)}{f^{(j)}(t)}
			= i\omega^{(j)} + \frac{\dot{g}^{(j)}(t)}{g^{(j)}(t)}.
		\end{equation}

	For the second equality above we just applied the definition of $f^{(j)}$. We now wish to explicitly separate the superoperator $\mathcal{L}^{(j)}_t = \dot{\Phi}^{(j)}_t\circ\left(\Phi^{(j)}_t\right)^{-1}$ from the operator $\varrho^{(j)}(t)$ on the right side of equation \eqref{eq:obtain_eqns:dot_rho_j_inv}, and then put it in Lindblad form with traceless jump operators. By inspection, we consider the action of the commutator $\left[\sigma_z,\cdot\right]$ on a generic $2\times2$ matrix $M=\sum_{m,n=0}^1 M_{mn}\kbb{m}{n}$. Simple algebra leads to 

		\begin{gather}	
		\left[ \sigma_z, M\right]=
			\begin{pmatrix}
				0 & -2M_{01} \\
				+2M_{10} & 0
			\end{pmatrix} \\
		\Leftrightarrow
		\begin{pmatrix}
			0 & M_{01} \\
			-M_{10} & 0
		\end{pmatrix}
		=-\frac{1}{2}
		\left[ \sigma_z, M\right].
		\label{eq:obtain_eqns:commut_z_action}
		\end{gather}

	Additionally, we take the elementary Lindblad-form dissipator generated by the jump operator $L=\sigma_z$ alone, 

		\begin{equation}
		\label{eq:obtain_eqns:dissip_z_def}
			\mathcal{D}_z\left[M\right] :=
			\sigma_z M \sigma_z\hc - \frac{1}{2}\left\lbrace \sigma_z\hc\sigma_z, M \right\rbrace ,
		\end{equation}
	obtaining 
	
		\begin{equation}
		\label{eq:obtain_eqns:dissip_z_action}
			\begin{pmatrix}
				0 & M_{01} \\
				M_{10} & 0
			\end{pmatrix}
			=
			-\frac{1}{2}
			\mathcal{D}_z\left[M\right].
		\end{equation}

	We now employ the real-imaginary decomposition $\varphi^{(j)}(t)=\varphi^{(j)}_R(t)+i\varphi^{(j)}_I(t)$ in \eqref{eq:obtain_eqns:dot_rho_j_inv}, which yields 
	
		\begin{equation}
		\label{eq:obtain_eqns:real_imag_master_passage}
		\begin{aligned}
			\mathcal{L}^{(j)}_t
			\left\lbrace 
			\varrho^{(j)}(t)
			\right\rbrace
			&=
			\dot{\varrho}^{(j)}(t) \\
			&= 
			\varphi^{(j)}_R(t)
			\begin{pmatrix}
				0 &
				\varrho^{(j)}_{01}(t)\\
				\varrho^{(j)}_{10}(t)
				& 0 
			\end{pmatrix}\\
			&+ i\varphi^{(j)}_I(t)
			\begin{pmatrix}
				0 &
				\varrho^{(j)}_{01}(t)\\
				-\varrho^{(j)}_{10}(t)
				& 0 
			\end{pmatrix}
		\end{aligned}
		\end{equation}
		
	We can finally apply the identities \eqref{eq:obtain_eqns:commut_z_action} and \eqref{eq:obtain_eqns:dissip_z_action} in \eqref{eq:obtain_eqns:real_imag_master_passage}, obtaining
		
		\begin{equation}
		\label{eq:obtain_eqns:master_eqn}
		\begin{aligned}
			\mathcal{L}^{(j)}_t
			\left\lbrace 
			\varrho^{(j)}(t)
			\right\rbrace
			=
			&-i \left[
			\frac{\varphi^{(j)}_I(t)}{2}\sigma_z^{(j)},
			\varrho^{(j)}(t)
			\right]\\
			&-\frac{\varphi^{(j)}_R(t)}{2}
			\mathcal{D}_z^{(j)}\left[
			\varrho^{(j)}(t)
			\right].
		\end{aligned}
		\end{equation}
	
	We have thus obtained the desired form \eqref{eq:canon_deph:exact_master_equation} with proper identifications (\ref{eq:canon_deph:effective_ham},\ref{eq:canon_deph:effective_omega},\ref{eq:canon_deph:dissip},\ref{eq:canon_deph:effective_gamma}). Note that in writing \eqref{eq:canon_deph:effective_omega} we used
	$\tilde{\omega}^{(j)}(t):=\varphi^{(j)}_I(t)=\Im\left[\varphi^{(j)}(t)\right] = \omega^{(j)} + \Im\left[ \dot{g}^{(j)}(t)/g^{(j)}(t)\right]$, where \eqref{eq:obtain_eqns:def_varphi} is employed, and similarly for $\gamma^{(j)}(t)$ in \eqref{eq:canon_deph:effective_gamma}. As a consistency check, one may notice that in the decoupling limit $\coupstr \to 0$ the functions $g^{(j)}(t)$ become constants:  $g^{(j)}(t) \to 1$, from the definitions \eqref{eq:canon_deph:def_g_ab}. Then \eqref{eq:canon_deph:effective_gamma} gives $\gamma(t) \to 0$ -- the dynamics is purely unitary -- and \eqref{eq:canon_deph:effective_omega} yields $\tilde{\omega}^{(j)}(t)\to \omega^{(j)}$ -- the bare frequencies are recovered.

\subsection{Existence of the inverse map}
\label{subsec:appx_obtain_eqns_exist}
	
	Recalling that the diagonal terms in \eqref{eq:canon_deph:local_states_sol} are constants, equation \eqref{eq:obtain_eqns:invert_01} shows explicitly that we can write $\varrho^{(j)}(0)$ as a function of $\varrho^{(j)}(t)$ -- i. e., invert the map $\Phi^{(j)}_t$ -- if and only if $f^{(j)}(t)\neq 0$, or, equivalently, $g^{(j)}(t)\neq 0$.\footnote{
		Indeed, by writing $\Phi^{(j)}_t$ in matrix form on a suitable basis, similar to the approach of \cite{Smirne_Vacchini_2010}, we can verify that $\det{\Phi^{(j)}_t} = \left|g^{(j)}(t)\right|^2$.
	} To study this condition we recall the definitions \eqref{eq:canon_deph:def_g_ab} and write them in compact notation, $g^{(j)}(t) =  p^{(\sim j)}_0 e^{-2i\coupstr t}+p^{(\sim j)}_1 e^{2i\coupstr t}$, as introduced in equation \eqref{eq:canon_deph:omega_eff_explicit_j}. We work out this expression,
		\begin{equation}
		\begin{aligned}
			g^{(j)}(t) &=  
			p^{(\sim j)}_0
			e^{-2i\coupstr t}+
			p^{(\sim j)}_1
			e^{2i\coupstr t}\\
			&=  
			\left( 1 - p^{(\sim j)}_1 \right)
			e^{-2i\coupstr t}+
			p^{(\sim j)}_1
			e^{2i\coupstr t}\\
			&= \cos{\left(2\coupstr t \right)} - 
				i \sin{\left(2\coupstr t \right)} + 
			2i p^{(\sim j)}_1 \sin{\left(2\coupstr t \right)}\\
			&= \cos{\left(2\coupstr t\right)} + i
			\left(  2p^{(\sim j)}_1 - 1\right) \sin{\left(2\coupstr t \right)}
		\end{aligned}
		\end{equation}
	In particular, one can see that
		
		\begin{equation}
		\label{eq:obtain_eqns:abs_gj_t}
			\left| g^{(j)}(t)\right|^2 =
			1 - 4p^{(\sim j)}_0p^{(\sim j)}_1 \sin^2\left(2\coupstr t\right).
		\end{equation}
	
	Therefore, the minimum value of $\left| g^{(j)}(t)\right|^2$, attained for $\sin^2\left(2\coupstr t\right)=1$, is 	
	
		\begin{equation}
		\begin{aligned}
			1 - 4p^{(\sim j)}_0p^{(\sim j)}_1 &= 
			\left(
				p_1^{(\sim j)}+
				p_0^{(\sim j)}
			\right)^2
			- 4p^{(\sim j)}_0p^{(\sim j)}_1 \\
			&= \left(
				p_1^{(\sim j)}-
				p_0^{(\sim j)}
			\right)^2 \\
			&= 4\left(
				p_1^{(\sim j)}-
				\frac{1}{2}
			\right)^2.
		\end{aligned}
		\end{equation}
	
	We conclude that $g^{(j)}(t) \neq 0$ -- i. e., $\Phi^{(j)}_t$ has an inverse map -- for any $t$ if $p_1^{(\sim j)}\neq 1/2$. In turn, if $p_1^{(\sim j)} = 1/2$, equation \eqref{eq:obtain_eqns:abs_gj_t} shows that $g^{(j)}(t)$ vanishes at the times $t^*$ such that  $\sin^2\left(2\coupstr t^*\right)=1$, and thus for these times, for this particular choice of initial state of the system ${\sim}j$, the dynamical map of the system $j$ is not invertible.

\bibliography{bibliography.bib}

\end{document}